\documentclass[modern]{aastex61}

\usepackage{amsfonts}
\usepackage{amsmath}
\usepackage{amssymb}
\usepackage{upgreek}
\usepackage{bm}
\usepackage{dcolumn}
\usepackage{epsfig}
\usepackage{graphicx}
\usepackage{placeins}
\usepackage{comment}
\usepackage{gensymb}
\usepackage{lineno}
\usepackage{wrapfig}
\usepackage{ragged2e}
\usepackage[utf8]{inputenc}
\usepackage{enumitem}

\usepackage{color}
\definecolor{spring}{rgb}{0.7,0.9,0.7}
\definecolor{brick}{rgb}{0.7,0.2,0.1}
\definecolor{redHL}{rgb}{1.0,0.5,0.5}
\definecolor{darkgreen}{rgb}{0.0,0.6,0.0}

\newcommand{\checkme}[1]{{#1}}

\newcommand{\Gaia}{\emph{Gaia}}
\newcommand{\Mc}{\mathcal{M}}
\usepackage[margin=1.0in]{geometry}

\begin{document}

\title{Astro2020 Decadal Science White Paper:\\Gravitational Wave Survey of Galactic Ultra Compact Binaries}

\author{Tyson B. Littenberg}
\affiliation{NASA Marshall Space Flight Center, Huntsville, AL 35812, USA}
\email{email: tyson.b.littenberg@nasa.gov}  

\author{Katelyn Breivik}
\affiliation{Canadian Institute for Theoretical Astrophysics, University of Toronto, 60 George Street, Toronto, Ontario M5S 3H8, Canada}

\author{Warren R. Brown}
\affiliation{Center for Astrophysics $|$ Harvard \& Smithsonian, 60 Garden St, Cambridge, MA 02139, USA}

\author{Michael Eracleous}
\affiliation{Department of Astronomy \& Astrophysics, The Pennsylvania State University, 525 Davey Lab, University Park, PA 16802, USA}

\author{J.~J.~Hermes}
\affiliation{Department of Physics and Astronomy, University of North Carolina, Chapel Hill, NC 27599, USA}
\affiliation{Hubble Fellow}

\author{Kelly Holley-Bockelmann}
\affiliation{Vanderbilt University, Nashville TN 37235, USA}

\author{Kyle Kremer}
\affiliation{CIERA, Northwestern University, Evanston, IL 60208, USA}

\author{Thomas Kupfer}
\affiliation{Kavli Institute for Theoretical Physics, University of California, Santa Barbara, CA 93106, USA} 
\affiliation{Department of Physics, University of California, Santa Barbara, CA 93106, USA}

\author{Shane L. Larson}
\affiliation{CIERA, Northwestern University, Evanston, IL 60208, USA}


\begin{abstract}
Ultra-compact binaries (UCBs) are systems containing compact or degenerate stars with orbital periods less than one hour.  Tens of millions of UCBs are predicted to exist within the Galaxy emitting gravitational waves (GWs) at mHz frequencies.  Combining GW searches with electromagnetic (EM) surveys like \textit{Gaia} and LSST will yield a comprehensive, multimessenger catalog of UCBs in the galaxy.  Joint EM and GW observations enable measurements of masses, radii, and orbital dynamics far beyond what can be achieved by independent EM or GW studies.  GW+EM surveys of UCBs in the galaxy will yield a trove of unique insights into the nature of white dwarfs, the formation of compact objects, dynamical interactions in binaries, and energetic, accretion-driven phenomena like Type Ia superonovae.   
\end{abstract}

\noindent \textbf{Thematic Areas:}\\
$\boxtimes$ Formation and Evolution of Compact Objects\\
$\boxtimes$ Multi-Messenger Astronomy and Astrophysics\\

\vskip 0.1in
\thispagestyle{empty}

\section{Introduction} \label{sec:intro}
\setcounter{page}{1}


The success of the Laser Interferometer Gravitational-wave Observatory has demonstrated the unique possibilities of gravitational wave (GW) astronomy. As ground-based GW observatories continue to reveal the gravitational universe in the kHz regime, 
%
%
extending the GW measurement window to mHz frequencies will reveal astrophysical sources much richer in number and variety.  Excitingly, this includes persistent sources that are readily observed ``electromagnetically'' with standard astronomical tools, but free of biases inherent in electromagnetic (EM) observations like interstellar extinction.

A cornerstone source-class in the GW frequency band between ${\sim}0.1$ and ${\sim}10$ mHz is the Galactic population of ultra-compact binaries (UCBs):  binary systems made up of two stellar-mass compact objects with orbital periods $<1$ hour.  Binaries are abundant in the Milky Way.  Extrapolating from known binaries of all orbital periods, $\mathcal{O}(10^7)$ are expected to be emitting GWs in the mHz band (e.g. \citet{Nelemans2001}), typically comprising two white dwarf (WD) stars.  Tens to hundreds of Galactic UCBs with black hole and/or neutron star components are also expected to be emitting GWs in the mHz band \citep{Nelemans2001, Lamberts2018}.
The majority of these sources will form and evolve as isolated binaries, however some UCBs may form dynamically in stellar clusters, in the Galactic center, or as members of triple systems \citep[e.g.,][]{Kremer2018,Banerjee2018,Antonini2017}.

The Laser Interferometer Space Antenna \citep[LISA;][]{LISA} mission has a GW measurement band between ${\sim}0.1$ and ${\sim}100$ mHz and is expected to individually resolve $\mathcal{O}(10^4)$ UCBs in the Galaxy \citep{Cornish2017}.
Because it is an all-sky all-time monitor, LISA will continuously track the UCBs' orbital evolution over its multi-year mission lifetime. 
UCBs are guaranteed multi-messenger systems, with \checkme{${\sim}1\%$} of galactic sources detected by LISA being localized to within 1 square degree in the first months of observing, and upwards of \checkme{${\sim}20\%$} after the first four years of the mission~\citep{Cornish2017}.  
A growing number of UCBs discovered by EM observations are known ``verificiation binary'' sources for LISA.  UCBs discovered by GW observations can be linked back to EM counterparts on the basis of both position and orbital period, i.e.\ using optical variability surveys.  




UCBs thus serve as multi-messenger laboratories.  Joint EM+GW observations provide physical constraints on masses, radii, and orbital dynamics far beyond what independent EM or GW observations can achieve alone \citep{Shah2014}.
Just as with the Hulse-Taylor binary~\citep{Hulse1975}, multi-messenger UCB systems can be studied well beyond the culmination of the GW discoveries, laying the foundation for decades of study utilizing UCBs as probes of relativistic and astrophysical processes.


\clearpage
\begin{wrapfigure}{r}{0.55\textwidth}
  \begin{center}
    \includegraphics[width=0.55\textwidth]{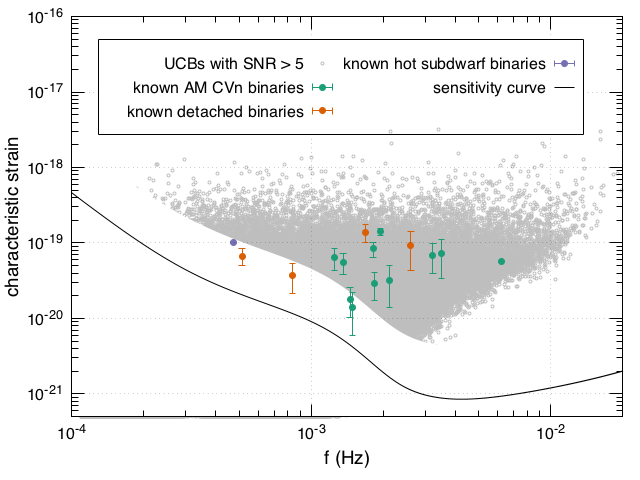}
  \end{center}
  \caption{\small Sensitivity plot for LISA assuming 4 years of observation showing the binaries which reach a $\rm{SNR} \sim 5$. Gray points are a simulated population, green circles are AM CVn systems, orange circles correspond to detached white dwarfs and the purple circles are the hot subdwarf binaries. Adapted from \citet{Kupfer2018}}
\label{fig:spectrum}
\end{wrapfigure}

\section{Gravitational-Wave Survey of Ultra-Compact Binaries} \label{sec:population}
Compared to compact object merger events, the orbital velocities, $v_{\rm orb}$, of the stars in mHz-band binaries are significantly less than the speed of light, $c$. As a result, the gravitational waveforms are comparatively simple to model. 
A subset of systems will also have a clearly measurable first time derivative of the frequency $\dot{f}$ and, after several years of observing, a small number of sources will also have a detectable second time derivative of the frequency $\ddot{f}$. Because UCBs are continuous GW sources, the signal-to-noise ratio (SNR) will improve over the observation time as $\sqrt{T}$.
Position and orientation information for the binaries comes from modulations imparted on the GW signal from the orbital motion of the detector, and long-duration observations enable monitoring of the frequency evolution of the binaries, which encodes valuable physics \citep[e.g. relativistic effects on the orbital motion, internal structure of WD stars, and mass transfer physics; see, for example][]{Taam1980,Savonije1986,Willems2008,Nelemans2010}. The combination of EM+GW measurements, then, will enable very sensitive tests of models for the evolution, mass transfer, and accretion in these systems.  

About $10^4$ individually resolvable UCBs are expected to be detected in the first year of a LISA-like mission \citep{Cornish2017}.  Population inferences made from the catalog of UCBs, such as the frequency and $\dot{f}$ distributions, will provide statistically robust insight into the complicated astrophysical processes undergone by binary stars, including the formation of the compact objects themselves, common envelope evolution, mass transfer, and the end state of these systems, perhaps as Type Ia supernovae, AM CVn systems, massive WDs, or subdwarf-O and R Corona Borealis stars \citep{Webbink1984}.  These same physical processes are at play to understand the formation channels of other compact binaries, including X-ray binaries \citep[e.g.,][]{vanHaaften2012} and the neutron star/black hole binary mergers observed by ground-based GW observatories \citep[e.g.,][]{Stevenson2015}.  Space-based GW observations will provide a long lever-arm on binary population synthesis models thanks to the enormous number of sources.

Because GWs propagate unobstructed through matter, UCBs will be detectable beyond the Galactic center and across the galaxy, whereas EM surveys are limited by intervening material in the Galactic plane.  Well-localized binaries 
will be used to infer the large scale structure of the Milky Way~\citep{Adams2012, Korol2018} perhaps reaching to nearby galaxies \citep{Korol2018b}.

The majority of GW sources in the Milky Way will not be individually resolvable in frequency space, but instead will blend together to form a source-confusion-limited astrophysical foreground which will be the dominant source of ``noise'' for LISA from \checkme{${\sim}0.4-3$ mHz}\footnote{e.g., see Fig. 9 of LISA Science Requirements Document ESA-L3-EST-SCI-RS-001}.  
The spatial distribution of these faint UCB sources follows that of the galaxy.  Because a GW detector's sensitivity depends on the orientation of the detector with respect to the GW sources, the confusion noise will vary in time.  The spectral shape of the confusion noise and the depth and shape of the amplitude modulations will provide additional insight into frequency and spatial distribution of UCBs.


\section{Ultra-Compact Binaries as Multi-Messenger Astrophysical Laboratories} \label{sec:counterparts}
Every UCB in the Galaxy with orbital period below $\sim$200 s will be clearly detected throughout the galaxy by LISA.
Thanks to their high SNR, these short-period UCBs will be identified early in a GW mission (within a few weeks of observing) and well localized, making them excellent candidates for multi-messenger observations.  These sources will also enable precision measurement of their orbital evolution.

The orbital evolution of two point particles, to leading post-Newtonian order (sufficient for UCBs because $v_{\rm orb}\ll c$), is completely determined by the orbital frequency and the chirp mass $\Mc\equiv (m_1 m_2)^{3/5}(m_1+m_2)^{-1/5}$, where $m_1$ and $m_2$ are the individual masses of the binary components. As discussed in Sec. \ref{sec:population}, $\Mc$ is not directly measured for typical UCBs, but is encoded in the GW amplitude $\mathcal{A}$ along with, $f$, inclination angle $\iota$, and the luminosity distance $D_L$.  For UCBs evolving only due to gravitational wave emission, $\dot{f} \propto \Mc^{5/3} f^{11/3}$, thus measuring $\dot{f}$ constrains $\Mc$, and the amplitude is used to determine the distance to the source (the polarization content of the GWs helps constrain the inclination).  


A large fraction of observed UCBs will have non-relativistic contributions to $\dot{f}$, which presents a new line of study.  Some of the most compact UCBs in the galaxy are helium mass-transfer AM CVn binaries in which $\dot{f}$ is dominated by the mass transfer physics between the two stars \citep{Kremer2017}. For such systems, an independent distance measurement (e.g., by \textit{Gaia}) decouples the frequency evolution into its different components \citep{Breivik2018}. The observation (or lack thereof) of AM CVns with a helium WD companion will place constraints the stability of mass transfer in WD UCBs and shed light on AM CVns as potential Type Ia supernovae progenitors \citep{Marsh2004, Sepinsky2014, Shen2015}. 
In detached WD UCBs, tidal theory predicts a $\sim$10\% enhancement to $\dot{f}$ because WDs tidally heat-up as they come into merger \citep{Benacquista2011, Piro2011, Fuller2013}.  It is also possible that many UCBs are members of hierarchical systems; \citet{Robson2018} show that a systematic change in $\dot{f}$ can constrain the orbit of triples with outer periods less than about 10 times the observation time baseline.  In all cases, long-term monitoring of $\dot{f}$ enables new constraints. 


\subsection{Known EM+GW sources}

There are 11 UCBs, already known from EM observations, that will be detected at SNR$\gtrsim$5 with LISA \citep{Kupfer2018}.  Most are helium mass-transfer AM CVn binaries consisting of a WD accretor and a helium donor star.  The highest GW amplitude system is HM Cnc with an orbital period of 321 s \citep{Roelofs2010}; this object will be detectable within weeks of observing with LISA.  The others are detached WD binaries; the highest GW amplitude system is SDSS J0651+28 with an orbital period of 765 s \citep{Brown2011}.  


EM observations are important to fully exploiting the GW observations.  Simply having an accurate EM sky position can improve measurement uncertainties from GW observations by a factor of two \citep{Shah2012}; adding EM constraints on binary inclination or orbital frequency change, i.e.\ from eclipse timing, ellipsoidal variations, or radial velocity measurements, further improves source characterization by a factor of 40 \citep{Shah2013}.  Similarly, combining the chirp mass obtained from GW observations and the mass ratio from optical spectroscopy radial velocity curves allows an independent measurement of the masses of the two components of the binary to exquisite precision \citep{Shah2014a}. 

Known UCBs are commonly single-lined spectroscopic binaries, in which the hottest object dominates the light of the system.  Radial velocity measurements from optical spectroscopy yield the ratio of masses in the binary given an inclination constraint.  EM time series photometry can constrain the binary inclination, the orbital period, the ratio of stellar radii, and $\dot{f}$ from eclipses, ellipsoidal variation, Doppler beaming, and other photometric signals commonly observed in known UCBs \citep[e.g.,][]{Hermes2012}.  EM astrometry, i.e.\ from {\it Gaia}, measures accurate position and parallax distance.  Thus systems with EM+GW observations make the best laboratories for UCB science because all of their fundamental properties can be measured extremely well by multiple methods. 

For example, tidal dissipation is expected to significantly influence physical conditions such as WD surface temperature and rotation rate prior to mass transfer or merger \citep{Fuller2012, Fuller2013}.  For the detached WD binary J0651+28, tidal dissipation should manifest itself as a $\sim$5\% increase in $\dot{f}$ over the GW-driven evolution~\citep{Piro2011, Benacquista2011}.  EM observers have measured J0651+28's $\dot{f}$ with 0.3\% accuracy \citep{Hermes2019}.  However, J0651+28's mass is not known well enough from EM measurements to test whether $\dot{f}$ significantly deviates from General Relativity.  GW measurements provide an independent measure of mass.  Combining EM and GW measurements can significantly improve our estimates of the system parameters \citep{Shah2014}, and the complementarity of the methods should enable a significant constraint on tidal heating in this merging pair of white dwarfs.

The number of EM UCBs detectable by a LISA-like mission will continue to grow over the coming years (e.g., see Sec.~\ref{sec:context} and ~\citet{Brown2017} for the recent discovery of a new multi-messenger candidate), though it is expected that a LISA-like mission will discover orders of magnitude more GW sources that are known to EM observations. 


\subsection{EM Follow-up of GW sources}

While dozens of UCBs known from their EM emission will be detectable by their GW emission, a larger number of UCBs will first be discovered by their GW emission \citep{Korol2017}.  Thousands of the UCB sources discovered by gravitational waves will be localized to within 1 square degree after two years of observations by a LISA-like mission \citep{Littenberg2012, Cornish2017}.  For all resolvable systems, space-based GW observatories will precisely measure the orbital period.  

The combination of GW sky position, orbital period, and possibly $\dot{f}$ provides a link to counterparts in EM variability surveys.  Approximately 10\% of $P_{\rm orb} = 10$~min, $(m_1,m_2)=(1.0,0.5)~{\rm M}_\odot$ double WD binaries will be eclipsing, given the ratio of WD radii to orbital separation.  Many more UCBs will exhibit photometric variability at an integer of orbital period, such as the reflection effect, Doppler beaming, or ellipsoidal variation observed in known UCBs \citep{Hermes2014}.  
Linking GW detections to EM light curves, EM astrometry, and EM spectroscopy will provide robust measurements of masses, radii, orbital separation, and inclination angle of UCBs beyond what can be achieved by either observing strategy on its own. 




\section{Contextualizing Ultra-Compact Binary Observations in 2030} \label{sec:context}
In the intervening years between now and when LISA begins operations, UCB science will continue to advance through the advent of powerful optical surveys.  Some of the most prolific instruments for UCB discoveries in the optical wavebands will be \Gaia~\citep{GAIA}, ATLAS \citep{atlas}, the Zwicky Transient Facility \citep[ZTF; ][]{ZTF}, BlackGEM \citep{blackgem}, the Large Synoptic Survey Telescope \citep[LSST;][]{LSST} and will be complemented with surveys in other frequency bands (e.g.\ eROSITA in the X-rays and perhaps upcoming UV missions). 

This will be complemented with the next generation of follow-up facilities like the 30m telescope or the James Webb telescope which will allow precise EM studies of UCBs.  The result is a ``bright'' future for UCB research poised for detailed EM+GW studies as soon as \emph{LISA} starts observing. 

Folding simulated Galactic populations through \Gaia\ and LSST response simulations, \citet{Korol2017} predict that these two EM surveys will discover a few hundred (\Gaia) to a thousand (LSST) UCBs that can be individually resolved with GWs.  UCBs are natural multi-messenger laboratories.
We conclude that a large number of UCBs, studied with GW+EM observations, will allow us to study a number of astrophysical phenomena that are of general importance to our understanding of the Universe, including accretion physics, high-energy phenomena like Type Ia supernovae, and the formation and evolution of compact objects.


\bibliographystyle{aasjournal}
\bibliography{master}

\end{document}